\documentstyle[prb,aps,psfig]{revtex}

\newcommand{\be}{\begin{equation}} 
\newcommand{\ee}{\end{equation}}
\newcommand{\bt}{\begin{tabular}} 
\newcommand{\et}{\end{tabular}}

\begin{document}

\draft

\twocolumn[\hsize\textwidth\columnwidth\hsize\csname@twocolumnfalse%
\endcsname

\title{Application of a minimum cost flow algorithm to the three-dimensional
gauge glass model with screening}

\author{Jens Kisker} \address{Institut f\"ur Theoretische Physik,
  Universit\"at zu K\"oln, D-50937 K\"oln, Germany}

\author{Heiko Rieger} \address{HLRZ, Forschungszentrum J\"ulich, 
  D-52425 J\"ulich, Germany}

\date{June 21, 1998}

\maketitle

\begin{abstract}
We study the three-dimensional gauge glass model in the limit of 
strong screening by using a minimum cost flow algorithm, enabling us to 
obtain {\em exact} ground states for systems of linear size $L\le48$. 
By calculating the domain-wall energy, we obtain the stiffness exponent 
$\theta = -0.95\pm 0.03$, indicating the absence of a finite temperature
phase transition, and the thermal exponent $\nu=1.05\pm0.03$.
We discuss the sensitivity of the ground state with respect to small 
perturbations of the disorder and determine the overlap length, 
which is characterized by the chaos exponent $\zeta=3.9\pm0.2$, 
implying strong chaos.
\end{abstract}

\pacs{74.40.+k, 74.60.-w, 64.60.Ak, 75.10.Nr}

] \newpage

In high-temperature superconductors, which are strongly type-II,
disorder plays an important role as a pinning mechanism for vortices.
Without disorder, as has been suggested by Abrikosov within mean-field
theory, flux lines form a triangular lattice. However, under the
influence of an external current perpendicular to the field, the
vortices will move since they experience a Lorentz force, and thus
energy is dissipated destroying the superconducting state. It has been
suggested \cite{Fisher,FFH} that disorder in the form of point defects
may pin the vortices at random positions leading to a superconducting
vortex glass phase, where the phase of the order parameter is random
in space, but frozen in time, similar to spin glasses.

A simplified model commonly used to study the vortex glass is the
gauge glass, which is believed to be in the same universality class
and contains the necessary prerequisites, disorder and frustration,
for glassy behavior.  Within this model, one neglects fluctuations in
the amplitude of the order parameter and only considers the phase of
the condensate. The Hamiltonian is given by
\begin{eqnarray}
H & = &-J\sum_{\langle i,j\rangle} \cos(\phi_i-\phi_j-A_{ij}-
\lambda_0^{-1} a_{ij}) \nonumber\\
& & +\frac{1}{2}\sum_\Box ({\bf \nabla}\times{\bf a})^2 \, ,
\label{gg_hamil}
\end{eqnarray}
where $\phi_i$ is the phase of the order parameter on site $i$ and $J$
is the interaction strength, henceforth set to $J=1$. The sum is over
all pairs $\langle i,j\rangle$ of nearest neighbors on a simple cubic
lattice of size $N=L^3$.  The quenched random vector potentials
$A_{ij}$ are drawn uniformly from the interval $[0,2\pi]$ and
represent the effect of disorder and an external magnetic field.
Screening of the interactions between vortices is incorporated by the
fluctuating vector potentials $a_{ij}$ which are integrated over from
$-\infty$ to $\infty$ under the gauge constraint ${\bf \nabla }\cdot
{\bf a}=0$, and $\lambda_0$ denotes the screening length. The limit
$\lambda_0\rightarrow0$ corresponds to strong screening, whereas
$\lambda_0\rightarrow\infty$ is the limiting case without screening.
The last term describes the magnetic energy, and is the sum over all
plaquettes of the lattice, where the curl is given as the directed sum
of the $a_{ij}$ around one plaquette.

Most of the theoretical work so far has concentrated on establishing
numerically the lower critical dimension of the gauge glass model,
both with and without screening of the interactions between vortices.
Without screening, there is no finite temperature transition to a
vortex glass phase in two dimensions \cite{FTY,Gingras,BY,KS},
whereas in three dimensions there is evidence for a finite $T_c$, as
has been found by domain wall renormalization group analyses (DWRG)
\cite{Gingras,BY,RTYF,KS,MG} and finite temperature Monte
Carlo simulations \cite{HS,WY1}, though due to limited system sizes
and insufficient statistics the earlier DWRG studies
\cite{Gingras,BY,RTYF} could not fully rule out that the lower
critical dimension is exactly $d=3$.

Sufficiently close to the critical point, screening effects become
important, since the correlation length $\xi$ diverges more strongly
than the screening length $\lambda$ and the two length scales
eventually become comparable \cite{BY}. The effect of screening was
investigated in Ref. \cite{BY} by a DWRG study and more recently in
Ref. \cite{WY} by means of a finite temperature Monte Carlo
simulation, and the results indicate that screening is a relevant
perturbation, destroying the finite temperature transition in three
dimensions, though the DWRG analysis could only be performed for
rather small system sizes $(L\le4)$.

In the present paper we reinvestigate the gauge glass model in the
limit of strong screening by performing a DWRG analysis using {\em
  exact} ground states, which we obtain via a minimum cost flow
algorithm from combinatorial optimization.  This algorithm allows us
to study systems with linear size up to $L=48$, which is considerably
larger than the system sizes in the previous studies \cite{BY,WY}.  In
addition, we study for the first time the sensitivity of the ground
state configurations with respect to small parameter changes, thereby
obtaining the {\em chaos} exponent.

We make use of the vortex representation of the gauge glass model,
which is obtained from the Hamiltonian in Eq. (\ref{gg_hamil}) by
making the Villain approximation \cite{Villain,JKKN,Kleinert}, which
replaces the exponentiated cosine term in the partition function by a
sum of periodic Gaussians, and then integrating out the spin wave
degrees of freedom. Thereby one obtains the vortex Hamiltonian
\cite{FTY}
\be
H_{V} = -\frac{1}{2}\sum_{i,j} ({\bf J}_i - {\bf b}_i) G(i-j) ({\bf J}_j - 
{\bf b}_j) \, ,
\label{vortex_hamil}
\ee
defined on the {\em dual} lattice, which again is a simple cubic
lattice.  The ${\bf J}_i$ are three-component integer variables
running from $-\infty$ to $\infty$ living on the links of the dual
lattice and satisfy the divergence constraint $({\bf \nabla} \cdot
{\bf J})_i = 0$ on every site $i$. The ${\bf b_i}$ are magnetic fields
which are constructed from the quenched vector potentials $A_{ij}$ by
a lattice curl, i.e. one obtains ${\bf b_i}$ as $1/(2\pi)$ times the
directed sum of the vector potentials on the plaquette surrounding the
link on the dual lattice ${\bf b_i}$ lives on. By definition, the
magnetic fields satisfy the divergence free condition $({\bf \nabla}
\cdot {\bf b})_i = 0$ on every site, since they stem from a lattice
curl. The vortex interaction is given by the lattice Green's function
\be
G(i,j)=J\frac{(2\pi)^2}{L^3}\sum_{{\bf k\neq 0}} \frac{1-\exp({\rm i}
{\bf k}\cdot ({\bf r}_i - {\bf r}_j))}{2\sum_{n=1}^d 
(1-\cos(k_n))+\lambda_0^{-2}}
\,.
\ee
In the case we are interested in, the strong screening limit
$\lambda_0\rightarrow0$, $G(i-j)$ reduces to $G(0) = 0$ for $i=j$ and
$G(i,j) = J(2\pi\lambda_0)^2$ for $i\neq j$ with exponentially small
corrections \cite{WY}. Thus if we subtract $J(2\pi\lambda_0)^2$ from
the interaction and measure the energy in units of
$J(2\pi\lambda_0)^2$, one obtains the simpler Hamiltonian
\be
H_V=\frac{1}{2}\sum_i ({\bf J}_i - {\bf b}_i)^2 \,.
\label{hamil}
\ee
We remark that $H_V$ is not trivial due to the divergence condition 
$(\nabla \cdot {\bf J})_i=0$.

Finding the ground state of the Hamiltonian in Eq. (\ref{hamil})
subject to the constraint $({\bf \nabla}\cdot {\bf J})_i =0$
can be interpreted as a {\em minimum cost flow problem} in the
language of combinatorial optimization. From this point of view, the 
problem can be restated as
\be
\quad \mbox{Minimize } \quad z({\bf J}) = \sum_i c_i({\bf J_i}) 
\label{minflow}
\ee
subject to the constraint $({\bf \nabla}\cdot {\bf J})_i =0$, where the
cost functions $c_i({\bf J_i}) = ({\bf J}_i -{\bf b}_i)^2/2$ have been 
defined.

The algorithm we shall use is the {\em successive shortest path
  algorithm} \cite{Ahuja,Rieger,RB}, which solves the problem in
polynomial time in this specific case. For the implementation we made
use of the LEDA programming library \cite{LEDA}.  We were able to
obtain {\em exact} ground states for system sizes up to $L=48$ on
ordinary workstations. The computation time increases approximately
with $N^2$, and one instance for $L=48$ took about 2.5 hours computer
time on a Sun Ultra2 (167MHz) workstation.  We remark that a similar
algorithm has already been used to calculate ground states of the
solid-on-solid model for a surface on a disordered substrate, which
also can be mapped on a minimum cost flow problem \cite{RB}.

We now present our results. First we address the question if the 
gauge glass model in the limit
of strong screening shows a finite temperature transition as 
presumably is the case
without screening, i.e. we investigate the stability of the ground state
with respect to thermal fluctuations. We make use of the concept of
domain wall renormalization, which has been applied to spin 
glasses in the same context \cite{BM}. The idea is to study the energy 
$\Delta E$ necessary to flip a cluster on length scale $L$. For long 
length scales $L$, one expects that 
\be
\Delta E \sim L^\theta \,,
\ee
where
$\theta$ is the stiffness exponent. The sign of $\theta$ now determines
whether there is a finite temperature phase transition.
If $\theta$ is positive, then the domain wall energy is 
increasing with cluster size, and one concludes that the ground state is 
stable with respect to thermal fluctuations.  
Correspondingly, if $\theta$ is negative, the argument is
that the ground state is unstable, since 
large cluster can be flipped by arbitrarily small energy. Thus in this case, 
the transition temperature is zero. 

The usual way to determine the defect- or domain wall energy is to
measure the energy difference between ground states obtained for
periodic and anti-periodic boundary conditions (bc), respectively. For
the model under consideration in the vortex representation
(\ref{hamil}), however, one has to incorporate a boundary term to
mimic the effect of a change in bc \cite{FTY,BY}.  This boundary term
is rather inconvenient to implement in the minimum cost flow algorithm
we are using, so we propose an alternative method to induce a domain
wall or elementary excitation. Note that the vortex Hamiltonian in
Eq. (\ref{hamil}) with periodic bc and without the additional boundary
term corresponds to fluctuating boundary conditions in the gauge glass
model \cite{Olsson,GTG}.  

In the gauge glass one changes from periodic to anti-periodic bc by
adding $\pi / L$ to the vector potentials in one spatial direction,
e.g.  $A_{ij}^x \rightarrow A_{ij}^x + \frac{\pi}{L}$. Such a shift of
the vector potential, however, has no effect on the Hamiltonian in the
vortex representation (\ref{hamil}), since the magnetic fields ${\bf
  b}_i$ are constructed from the vector potential by a lattice curl
and thus remain unchanged. Only the argument of the boundary term
changes, which can be compensated by forming one or several vortex
loops with total area \cite{FTY,BY} $L^2/2$. From this observation we derive our basic idea: First we consider the
vortex Hamiltonian Eq.  (\ref{hamil}) with periodic bc and calculate
the exact ground state configuration $\{ {\bf J}^0\}$. The energy
$E_0(\{ {\bf J}^0\})$ of this state is obtained via $H_V$ in Eq.
(\ref{hamil}).  We then determine the global flux $f$ of this
configuration in one spatial direction, e.g. in the x-direction
\be
f_x = \frac{1}{L} \sum_i J_i^x \,,
\ee
where $f_x$ can be interpreted as a total winding number. Next, we
gradually decrease all costs in the x-direction of the minimum flow
problem in Eq. (\ref{minflow}) simultaneously, which makes global flux
in this direction energetically more favorable, whereas the cost for
all topologically simply connected loops (those with winding
number zero around the 3d torus) remains unchanged.  We reduce the
costs until we obtain a configuration $\{ {\bf J}^1\}$ with global
flux $f_x+1$, which is an elementary low energy excitation with
length scale $L$.  We calculate the energy $E_1$ of this configuration
again with $H_V$ in Eq. (\ref{hamil}) with the {\em same} magnetic
fields ${\bf b}_i$ as before. The domain wall energy is then given by
$\Delta E = E_1 - E_0$, which is always positive since the new
state $\{ {\bf J}^1\}$ with flux $f_x+1$ corresponds to an excited
state.

In a small fraction $(\approx 5\%)$ of the samples, the flux changes
discontinuously by more than one unit upon slowly decreasing the costs
as described above. However, the resulting configuration still
represents an elementary excitation of length scale $L$ and we can
also use this configuration for calculating the energy.

We note that it is easy to see that in the vortex (or Villain)
representation of the pure classical three-dimensional XY-model which
is given by $H=\sum_i {\bf J}_i^2$, the elementary low energy
excitation is indeed given by a configuration with one additional flux
line of length $L$ winding once around the 3d torus. The energy
difference between the ground state ($f_x=0$) and the excited state
($f_x=1$) then simply is $\Delta E \sim L$, which corresponds to the
energy of a spin wave excitation with minimum wave vector in the 3d
XY-model in the phase representation.

\begin{figure}
\begin{center}
\leavevmode
\psfig{file=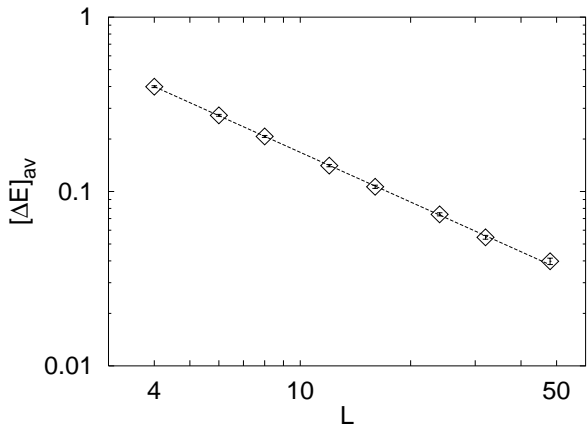,width=\columnwidth}
\end{center}
\caption{The domain wall energy $[\Delta E]_{{\rm av}}$ in a log-log plot. The
straight line is a fit to $[\Delta E]_{{\rm av}} \sim L^\theta$ with
$\theta=-0.95\pm0.03$. This implies a thermal exponent of 
$\nu=1.05\pm0.03$.
The disorder average is over 500 samples for $L=48$, 1500 samples for $L=32$,
and for the smaller sizes several thousand samples have been used.}
\label{fig1}
\end{figure}

Fig. \ref{fig1} shows the domain wall energy $\Delta E$ vs. $L$ for $L\le 48$
in a double logarithmic plot. One observes a straight line behavior which 
can be nicely fitted by
\be
\Delta E \sim L^{\theta} \quad \mbox{with} \quad \theta = -0.95 \pm 0.03
\,.
\label{def_en}
\ee

Thus we reestablish that $T_c=0$ for the gauge glass model in the
strong screening limit, as has been found in Refs. \cite{BY,WY}. From
the stiffness exponent $\theta$ the thermal exponent $\nu$, which
describes the divergence of the correlation length, can be calculated.
For $T_c=0$ it is $\xi \sim T^{-\nu}$ and by equating the thermal
energy with the energy of a low lying excitation on the length scale
of the correlation length, it follows that
\be
\nu =\frac{1}{|\theta|}\,.
\ee
From this relation we obtain $\nu=1.05\pm0.03$, which agrees well with
a result from a finite temperature Monte Carlo simulation for the same
model by Wengel and Young \cite{WY}, who were able to study system
sizes $L\le 12$ and found a zero temperature phase transition with
$\nu = 1.05\pm 0.1$.

Next we want to discuss the issue of {\em chaos} in the gauge glass
model.  From spin glasses it is known, that infinitesimal changes of
parameters like the temperature or the couplings can have quite a
dramatic effect on the ground state or equilibrium configuration
\cite{BM,RSBDJ}. It is argued that a so called overlap length $L^*$
exists, which is a measure for the length scale, up to which the
domain structure essentially remains unchanged after an infinitesimal
perturbation of a parameter. For length scales larger than the overlap
length, the domain structure changes.  For spin glasses, the overlap
length is given by \cite{BM}
\be
L^* \sim\delta^{-\frac{1}{\zeta}} \quad \mbox{with} \quad 
\zeta = \frac{d_s}{2} - \theta \,,
\label{ov_length}
\ee
where the result is obtained by equating the energy of a droplet
excitation $L^\theta$ and the energy change caused by the perturbation
which is proportional to $L^{d_s/2}$, where $d_s$ is the fractal
dimension of the droplets.

To pursue a similar investigation in the gauge glass model we study
the resulting changes in the ground state configuration when the
random vector potentials $A_{ij}$ are perturbed by a small amount.  To
be specific, we define new vector potentials by $A_{ij}' = A_{ij}
+\epsilon_{ij}$, where $\epsilon_{ij}$ is randomly drawn from the
interval $[-\delta,\delta]$.  We calculate the ground state for both
realizations of the disorder $\{A_{ij}\}$ and $\{A'_{ij} \}$, and
define the distance $D$ between the resulting ground state configurations
$\{{\bf J}\}$ and $\{{\bf J'}\}$ by
\be
D_\delta = \sum_i ({\bf J}_i -{\bf J'}_i)^2\,.
\label{overlap}
\ee
Note that for small values of $D_\delta(L)$ the ground states are more
correlated than for larger values.

To determine the chaos exponent, we perform a scaling plot of the data
for the overlap function $D_\delta(L)$. Guided by Eq. (\ref{ov_length}),
we attempt a scaling plot with the scaling variable
$L\delta^{1/\zeta}$, where $\zeta$ is the chaos exponent.  Such a plot
is shown in Fig. \ref{fig2}, and one observes that the data scales
nicely with $\zeta=3.9\pm0.2$.  This relatively large value of $\zeta$
implies {\em strong} chaos, since in the limit of a vanishing
perturbation $\delta \rightarrow 0$, the overlap length, which is
proportional to $\delta^{-1/\zeta}$, increases slowly.  The value of
$\zeta=3.9\pm0.2$ is considerably larger than for instance the one for
the two-dimensional Ising spin glass, where $\zeta=0.95\pm0.05$ has
been obtained \cite{RSBDJ,NY}.

\begin{figure}
\begin{center}
\leavevmode
\psfig{file=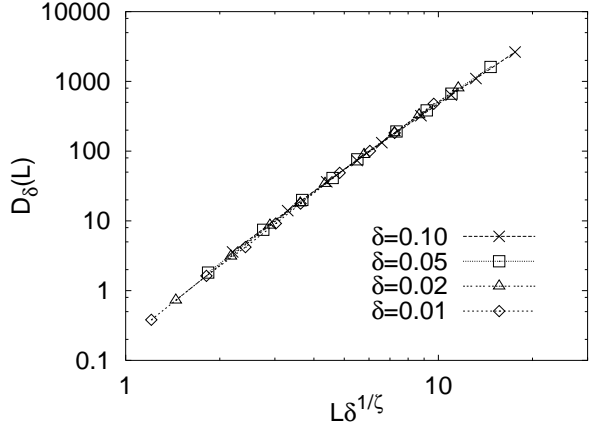,width=\columnwidth}
\end{center}
\caption{Scaling plot of the overlap function $D_\delta(L)$. The data scale
  nicely with the scaling variable $L\delta^{1/\zeta}$. The chaos
  exponent is given by $\zeta=3.9\pm0.2$. The system sizes studied
  range from $L=4$ to $L=32$, where 5000 samples have been used for
  $L\le20$, 2000 samples for $L=24$ and 500 samples for $L=32$.}
\label{fig2}
\end{figure}

Chaos cannot only be observed with respect to perturbations of the
disorder, but also with respect to small temperature changes
\cite{BM}. For a continuous bond distribution and low temperatures, the
overlap length is expected \cite{BM} to behave as $L^*_{th} \sim
T^{-2/\zeta}$ for $T_c=0$. The other relevant length scale is the
correlation length $\xi\sim\delta^{-\nu}$, and for the 2d Ising spin
glass with $\nu\approx2$ and $2/\zeta\approx2$, the exponents
characterizing the two length scales $L^*_{th}$ and $\xi$ appear to be
equal \cite{RSBDJ,NY,KHS}, so one might speculate that they are
related. However, for the gauge glass model we find
$2/\zeta\approx0.5$ and $\nu\approx1$, i.e.  the divergence of the
thermal correlation length is much stronger. It would be interesting
to study chaos with respect to temperature changes in the gauge glass
explicitly.

Summarizing we studied the strongly screened gauge glass model in the
vortex representation. Using a polynomial time minimum cost flow
algorithm we could deal with much larger system sizes than considered
before in the literature. We calculated the {\em exact} ground states
and also configurations with a (global) low energy excitation. In this
way we could perform a finite size scaling analysis of the so called
domain wall energy and obtained a pretty accurate estimate for the
stiffness exponent that is $\theta=-0.95\pm0.03$.  From this we can
draw two conclusions: a) since it is clearly negative there is no
superconducting glass phase (or vortex glass phase) at non-vanishing
temperature, and b) the thermal correlation length diverges only with
$T\to0$ as $\xi_{\rm th}\sim T^{-\nu}$ with an exponent
$\nu\sim1.05\pm0.03$, which is in agreement with Refs.\ \cite{BY,WY}.
Finally, we studied the effect of a small perturbation of the disorder
on the ground state domain structure and found that the overlap length
is characterized by the chaos exponent $\zeta=3.9\pm0.2$. This is a
pretty large value implying a fast destruction of ground state
correlations by thermal fluctuations.

%\section*{Acknowledgments}
We thank U.\ Blasum and Y.\ Dinitz for very inspiring discussions.
This work was supported by the Deutsche Forschungsgemeinschaft (DFG)
and was performed within the Sonderforschungsbereich (SFB) 341.


\begin{references}

\bibitem{Fisher}
  \vskip-1cm
  M.~P.~A.~Fisher, Phys. Rev. Lett. {\bf 62}, 1415 (1989).

\bibitem{FFH}
  D.~S.~Fisher, M.~P.~A.~Fisher and D.~A.~Huse, 
  Phys. Rev. B {\bf 43}, 130 (1991).

\bibitem{FTY}
  M.~P.~A. Fisher, T.~A. Tokuyasu and A.~P. Young, 
  Phys. Rev. Lett. {\bf 66}, 2931 (1991).

\bibitem{Gingras}
  M.~J.~P.~Gingras,
  Phys. Rev. B {\bf 45}, 7547 (1992).

\bibitem{BY}
  H.~S.~Bokil and A.~P.~Young, 
  Phys. Rev. Lett. {\bf 74}, 3021 (1995).

\bibitem{KS}
  J.~M.~Kosterlitz and M.~V.~Simkin, 
  Phys. Rev. Lett. {\bf 79}, 1098 (1997).

\bibitem{RTYF}
  J~.D.~Reger, T.~A.~Tokuyasu, A.~P.~Young and M.~P.~A.~Fisher,
  Phys. Rev. B {\bf 44}, 7147 (1991).

\bibitem{MG}
  J.~Maucourt and D.~R.Grempel,
  cond-mat/9802242.

\bibitem{HS}
  D.~A.~Huse and H.~S.~Seung,
  Phys. Rev. B {\bf 42}, 1059 (1990).

\bibitem{WY1}
  C.~Wengel and A.~P.~Young,
  Phys. Rev. B {\bf 56}, 5918 (1997).

\bibitem{WY}
  C.~Wengel and A.~P. Young, 
  Phys. Rev. B {\bf 54}, R6869 (1996).

\bibitem{Villain}
  J.~Villain,
  J. Physique {\bf 36}, 581 (1975).

\bibitem{JKKN}
  J.~V.~ Jos\'e, L.~P.~Kadanoff, S.~K.~Kirkpatrick, and D.~R. Nelson,
  Phys. Rev. B {\bf 16}, 1217 (1977).

\bibitem{Kleinert}
  H.~Kleinert, {\em Gauge fields in Condensed Matter}, (World Scientific, 
  Singapore, 1989).

\bibitem{Ahuja}
 R.~Ahuja, T.~Magnanti and J.~Orlin, 
 {\em Network Flows}, (Prentice Hall, New Jersey, 1993).

\bibitem{Rieger}
  H.~Rieger, {\em Frustrated Systems: Ground State Properties via
    Combinatorial Optimization},
  Lecture Notes in Physics 501 (Springer Verlag Heidelberg, 1998).

\bibitem{RB}
  H.~Rieger and U.~Blasum,
  Phys. Rev. B {\bf 55}, R7394 (1997);   
  U.~Blasum, W.~ Hochst\"attler, U.~Moll and H.~Rieger,
  J. Phys. A {\bf 29}, L459 (1996).


\bibitem{LEDA}
  Information on the LEDA (Library of efficient data types and algorithms)
  library can be found at http://www.mpi-sb.mpg.de/LEDA/leda.html.

\bibitem{BM}
  A.~J.~Bray and M.~A.~Moore, Phys. Rev. Lett. {\bf 58}, 57 (1987).

\bibitem{Olsson}
  P.~Olsson, Phys. Rev. B {\bf 52}, 4511 (1995).

\bibitem{GTG}
  P.~G.~ Gupta, S.~Teitel and M.~J.~P.~Gingras,
  Phys. Rev. Lett. {\bf 80}, 105 (1998).

\bibitem{RSBDJ}
  H.~Rieger, L.~Santen, U.~Blasum, M.~Diehl and M.~J\"unger,
  J. Phys. A {\bf 29}, 3939 (1996).

\bibitem{NY}
  M.~Ney-Nifle and A.~P.~Young,
  J. Phys. A {\bf 30}, 5311 (1997).

\bibitem{KHS}
  N.~Kawashima, N.~Hatano and M.~Suzuki,
  J. Phys. A {\bf 25}, 4985 (1992).

\end{references}
\end{document}